\documentclass[preprint]{aastex}
\usepackage{psfig}

\newcommand{\Ha}{H$\alpha$}
\newcommand{\Hb}{H$\beta$}
\newcommand{\HII}{\ion{H}{2}}
\newcommand{\OI}{[\ion{O}{1}]}
\newcommand{\OII}{[\ion{O}{2}]}
\newcommand{\OIII}{[\ion{O}{3}]}
\newcommand{\NII}{[\ion{N}{2}]}
\newcommand{\SII}{[\ion{S}{2}]}
\newcommand{\NeIII}{[\ion{Ne}{3}]}
\newcommand{\CaII}{\ion{Ca}{2}}
\newcommand{\MgIb}{\ion{Mg}{1}~$b$}

\newcommand{\gtwid}{\mathrel{\raise.25ex\hbox{$>$\kern-.80em\lower1ex\hbox{$\sim$}}}}

\shorttitle{``HIDDEN'' SEYFERT 2 GALAXIES}
\shortauthors{MORAN ET AL.}

\begin{document}

\title{``Hidden'' Seyfert 2 Galaxies and the X-ray Background}

\author{Edward C.\ Moran}
\affil{Astronomy Department, Wesleyan University, Middletown, CT 06459
       \\ \smallskip{\rm and}\\}

\author{Alexei V.\ Filippenko and Ryan Chornock}
\affil{Department of Astronomy, University of California,
                 Berkeley, CA 94720-3411}

\begin{abstract}
Obscured active galactic nuclei, which are classified optically as type~2
(narrow-line) Seyfert galaxies in the local universe, are by far the most
promising candidates for the origin of the hard (2--10 keV) X-ray background
radiation.  However, optical follow-up observations of faint X-ray sources
in deep {\it Chandra\/} images have revealed surprising numbers of apparently
normal galaxies at modest redshift.  Such objects represent $\sim$~40--60\%
of the sources classified in deep {\it Chandra\/} surveys, raising the
possibility that the X-ray galaxy population has evolved with cosmic time.
Alternatively, most of the faint X-ray galaxies in question are so distant
that their angular diameters are comparable to the slit widths used in
ground-based spectroscopic observations; thus, their nuclear spectral
features may be overwhelmed (``hidden'') by host-galaxy light.  To test
this hypothesis, we have obtained integrated spectra of a sample of nearby,
well-studied Seyfert 2 galaxies.  The data, which accurately simulate
observations of distant {\it Chandra\/} sources, demonstrate convincingly
that the defining spectral signatures of Seyfert 2s {\it can\/} be hidden
by light from their host galaxies.  In fact, 60\% of the observed objects
would not be classified as Seyfert 2s on the basis of their integrated
spectra, similar to the fraction of faint X-ray sources identified with
``normal'' galaxies.  Thus, the numbers of narrow-line active galaxies in
deep {\it Chandra\/} surveys (and perhaps all ground-based spectroscopic
surveys of distant galaxies) are likely to have been underestimated.
\end{abstract}

\newpage
\section{Introduction}

Identification of the classes of objects responsible for the cosmic X-ray
background radiation (XRB) has been one of the fundamental pursuits of X-ray
astronomy for the past several decades.  It now appears that different
populations produce the majority of the background in different X-ray energy
bands.  In the soft 0.5--2 keV range, active galactic nuclei (AGNs) with broad
optical emission lines (i.e., Seyfert~1 galaxies and quasars) are the dominant
contributors to the XRB (Schmidt et al.\ 1998).  These objects, however,
have steep X-ray spectra and cannot explain the flat slope of the XRB spectrum
at higher energies.  Type~2 (narrow-line) Seyfert galaxies have thus emerged
as the most promising candidates for the origin of the hard (2--10 keV) XRB.
The soft X-ray fluxes of these objects are heavily absorbed by dense
circumnuclear gas, presumably the same material that obscures their broad
emission-line regions.  Models (Madau, Ghisellini, \& Fabian 1994; Comastri
et al.\ 1995; Gilli, Salvati, \& Hasinger 2001) and X-ray observations of
nearby sources (Moran et al.\ 2001) have demonstrated that Seyfert~2 galaxies
have both the space density and X-ray spectral properties necessary to account
for the intensity and spectrum of the hard XRB.

The {\it Chandra X-ray Observatory}, which resolves the majority of the XRB
in deep exposures, provides an opportunity to confirm the Seyfert~2 hypothesis
directly.  Surprisingly, however, follow-up optical observations of faint
sources in the deepest {\it Chandra\/} images have revealed a significant
population of apparently {\it normal\/} galaxies and far fewer Seyfert 2s
than expected (e.g., Mushotzky et al.\ 2000; Barger et al.\ 2001a, 2001b,
2002; Hornschemeier et al.\ 2001).  ``Normal'' galaxies represent $\sim$
40--60\% of the hard X-ray sources classified in deep surveys; many exhibit
evidence of absorption in the X-ray band, but emission lines are very weak
in their starlight-dominated optical spectra.  The multiwavelength properties
of the prototypical example of this class are described by
Comastri et al.\ (2002a).  The prevailing interpretation is that optically
normal, X-ray--luminous galaxies are common at modest redshifts, perhaps
reflecting a tendency for AGNs in the past to be more absorbed or have weaker
ionizing continua than present-day Seyfert galaxies (Barger et al.\ 2001a).
This hints at the possibility that the X-ray galaxy population has evolved
with cosmic~time.

Alternatively, we propose that the results of the deep {\it Chandra\/} surveys
may be heavily influenced by the limitations of ground-based observing.  In
spectroscopic observations of nearby AGNs, light is collected through a small
aperture centered on the nucleus, which excludes most of the starlight from
the host galaxy.  The faint sources that produce the hard XRB, in contrast,
are very distant (the majority are at redshifts $z \gtwid 0.5$); thus, when
they are observed, the entire galaxy (or a large fraction of it) falls within
the spectrograph slit.  (At $z \approx 1$, 10 kpc corresponds to
$\sim$~1\farcs5.)  Combined with the low spectral resolution that has been
employed to date ($\sim$ 12--20 \AA ) and the modest signal-to-noise ratios
frequently obtained (the optical counterparts are generally quite faint),
the additional galaxy light (from stars and \HII\ regions) could lead
to the appearance that some distant {\it Chandra\/} sources are associated
with normal galaxies rather than with Seyfert~2s.

To investigate this possibility we have obtained {\it integrated\/} spectra
of nearby Seyfert~2 galaxies that are known to be absorbed X-ray sources.
Our techniques simulate ground-based spectroscopic observations of distant
X-ray galaxies, allowing us to evaluate whether the emission-line signatures
of their activity can be overwhelmed (or ``hidden'') in spectra of their
integrated light.  This project draws inspiration from a series of papers
by Kennicutt (1992a, 1992b), who obtained integrated spectra of nearby
galaxies of all types to assist ``the classification of unresolved galaxies
at large distances [and] the spectroscopic identification of extended sources
detected in infrared, radio, and X-ray surveys'' (Kennicutt 1992a).  Our
similarly motivated work focuses on the class of objects believed to be
responsible for the hard X-ray background.

\section{Observations and Analysis}
The objects for this study are drawn from the distance-limited sample of
Seyfert galaxies compiled by Ulvestad \& Wilson (1989).  There are 31
confirmed Seyfert~2s in the sample (Moran et al.\ 2000), almost all of
which have been observed in the 1--10 keV band with the {\sl ASCA\/}
satellite (Moran et al.\ 2001).  The Ulvestad \& Wilson sample is ideal
for this project for several reasons.  First, because of its distance-limited
nature, the sample is relatively unbiased.  Second, the hard X-ray
luminosities of the objects ($\sim$~$10^{41}$--$10^{43}$ ergs~s$^{-1}$)
are comparable to those of the ``normal'' galaxies
in deep {\it Chandra\/} surveys (Barger et al.\ 2001a, 2001b; Hornschemeier
et al.\ 2001).  And finally, because all of the objects are located within
$\sim$~60~Mpc, most are close enough that we can isolate the emission from
their nuclei yet distant enough (i.e., with sufficiently small angular
diameters) that we can measure their integrated spectra with relative ease.

Our data were acquired with the 3-m Shane reflector at Lick Observatory
over the course of three runs in August and November 2001 and February 2002.
We employed 600 line mm$^{-1}$ dispersing elements on each arm of the Kast
double spectrograph (Miller \& Stone 1993), which provided coverage of the
3400--8100 \AA\ range. For each galaxy, we obtained a 300~s exposure of the
nucleus using a $2''$ slit oriented at the parallactic angle (Filippenko
1982).  The nuclear spectra represent $4''$--$5''$ extractions along the
slit and have a resolution of 5--6~\AA\ (FWHM).
Integrated spectra of each object were obtained by scanning
an east-west slit across the galaxy in declination at a constant rate.
Six scans were obtained for most galaxies, although a few of the brighter
objects were scanned three to five times.  For the scans, the width of
the slit was increased to $6''$ to improve our observing efficiency and
to degrade the spectral resolution to 15--16~\AA , comparable to that
used in observations of faint {\it Chandra\/} sources (Barger et al.\ 
2002; Hornschemeier et al.\ 2001).  The angular distances of the scans
($30''$--$120''$ in declination) were dictated by the diameters of the
galaxies in high-contrast images from the Digitized Sky Survey (DSS).
Each scan had a duration of 600~s; combined with the slit width and scan
distances, this yielded effective exposures of 30~s to 120~s per scan
for each object's integrated spectrum.  These spectra include most of the
light from each galaxy above the sky background (i.e., $\sim 35''$ to
$\sim 90''$ extractions).

We observed a total of 18 objects from our sample --- all those at $\delta
> -23^{\circ}$ with projected angular sizes in right ascension (again,
estimated from DSS images) of less than $2'$.  The latter restriction
ensured that we would be able to perform accurate sky background subtraction
with light that entered the slit near the two ends. The galaxies observed
and some pertinent data are listed in Table~1.  The optical and X-ray
properties of the 13 objects we did not observe do not differ significantly.

\section{Integrated Spectra of Seyfert 2 Galaxies}
Recently published spectroscopic studies of faint {\it Chandra\/} sources are
based mainly on data acquired with 8--10~m telescopes (Barger et al.\ 2002;
Hornschemeier et al.\ 2001; Tozzi et al.\ 2001).  In most respects, the
integrated spectra we have obtained accurately simulate observations of
distant type~2 AGNs with these instruments.  The scan distances and
extractions employed for our sample include most of the light from the
galaxies; they correspond to physical sizes of $\sim$ 8--20 kpc, similar
to that which is typically included within a 1.5--$2''$ slit for a distant
{\it Chandra\/} source.  Our spectral resolution, as discussed above, is
comparable to that used in {\it Chandra\/} source surveys.  In addition,
the effective exposure times associated with our scans are equivalent to
1.5--2~hr Keck exposures of objects $\sim$ 7--8 mag fainter, i.e., objects
with visual magnitudes in the $m \approx$ 20--22 range.  These roughly
correspond to the Keck exposure times and magnitudes of distant
{\it Chandra\/} sources that have been classified spectroscopically
(Barger et al.\ 2001a, 2001b, 2002).

Examples of our results are displayed in Figure~1.  The nuclear spectrum of
NGC 262 (= Mrk 348; lower trace of Fig.~1a) bears all the characteristics
of a type~2 Seyfert galaxy: strong \OII\ $\lambda 3727$, \NeIII\ $\lambda
3869$, \Hb\ $\lambda 4861$, \OIII\ $\lambda\lambda 4959,5007$, \OI\ $\lambda
6300$, \Ha\ $\lambda 6563$, \NII\ $\lambda 6584$, and \SII\ $\lambda\lambda
6717,6731$ emission lines with high \OIII /\Hb , \OI /\Ha , \NII /\Ha , and
\SII /\Ha\ flux ratios (Veilleux \& Osterbrock 1987). These same features
are clearly present in the integrated spectrum of the galaxy (upper trace
of Fig.~1a), despite the presence of a significant amount of additional
starlight.  Thus, NGC~262 retains its Seyfert~2 classification in the
integrated spectrum.

The spectra in Figures 1b--e tell a much different story.  Once again, the
nuclear spectra of these galaxies exhibit Seyfert~2 emission-line signatures.
In the integrated spectra, however, the lines are far less prominent, and
some are completely obliterated.  The strongest spectral features are the
\CaII\ $\lambda\lambda 3934,3968$, G band ($\lambda 4300$), \MgIb\ $\lambda
5176$, and Na~I D $\lambda 5893$ stellar absorption lines typically observed
in the spectra of normal early-type galaxies.  Note that the \NeIII\ $\lambda
3869$ emission line, which has been used in {\it Chandra\/} source surveys
to identify AGNs (Barger et al.\ 2001a, 2001b; Hornschemeier et al.\ 2001),
is not visible in the integrated spectra.  Thus, \NeIII\ emission cannot be
relied upon to indicate the presence of an active nucleus.

The data for NGC~3982 (Fig.~1f) provide an interesting third alternative.
Although the emission lines in both the nuclear and integrated spectra of
this object are strong, the line-flux ratios in the two differ dramatically.
For example, in the nuclear spectrum, \NII\ $\lambda 6584$/\Ha\ $\approx$ 1
and \OIII\ $\lambda 5007$/\Hb\ $\approx$ 16, which is typical for a
Seyfert~2.  In the integrated spectrum, however, \NII /\Ha\ = 0.37 and
\OIII /\Hb\ = 0.82, consistent with the values observed in \HII\ regions
(e.g., Veilleux \& Osterbrock 1987).  The \OI /\Ha\ and \SII /\Ha\ ratios
also resemble those of \HII\ regions.  Given only the integrated spectrum
of NGC 3982, we would classify it as a starburst galaxy rather than a type~2
Seyfert!  We obtain a similar result for another object in our sample, NGC
1667.  Thus, it is possible for the emission from star-forming regions to
drown out an active nucleus in a galaxy, which sheds light on the nature of
the ``composite'' starburst/Seyfert galaxies uncovered during optical surveys
of {\sl ROSAT\/} sources (Moran, Halpern, \& Helfand 1996).

\section{Discussion and Conclusions}
The last column of Table 1 indicates our assessment of the classifications
of our galaxies based on their integrated spectra.  While seven objects with
strong lines in the nucleus remain as Seyfert~2s in the integrated spectra,
11 others (or $\sim$~60\%) would {\it not\/} be classified as AGNs.  This
is comparable to the fraction of {\it Chandra\/} sources ($\sim$~40--60\%)
identified with ``normal'' galaxies in deep surveys (Barger et al.\ 2001a,
2001b, 2002; Hornschemeier et al.\ 2001).  Among the normal-looking subset
of our sample, nine objects resemble early-type galaxies with red continua
and strong absorption features, and two objects have spectra reminiscent of
starburst (or post-starburst) galaxies.

All of the objects we observed display some evidence of emission lines in
their integrated spectra.  While the same is true of all of the ``normal''
galaxies whose spectra are presented by Barger et al.\ (2001a, 2001b) and
Hornschemeier et al.\ (2001), 19\% of the sources classified by Barger et
al.\ (2002) are described as ``absorption-line'' galaxies.  Some of these
objects may in fact have faint emission lines --- weak \OII\ and \OIII\
emission is visible in the average absorption-line galaxy spectrum shown
in Figure~5h of Barger et al.\ (2002).  Others, however, may be devoid
of emission features in the data obtained.  Two factors are likely to
contribute to the relatively high incidence of absorption-line systems in
the {\it Chandra\/} surveys:

\noindent
(1) The \Ha\ and \NII\ emission lines can be the strongest lines in the
spectra of Seyfert galaxies; as such, they are vital for accurate spectroscopic
classifications.  In fact, the classifications of two of the persistent
Seyfert~2s in our sample (Mrk~1066 and MCG $+$01-27-020) would be ambiguous
without information about the lines near \Ha .  However, $\sim$~70\% of the
spectra of supposedly normal galaxies published by Barger et al.\ and
Hornschemeier et al.\ do not cover the \Ha\ region because of the high
redshifts of the objects.  It is possible that some of the absorption-line
galaxies would exhibit emission lines if this region were observed.

\noindent
(2) Our relatively small sample of nearby objects does not include all possible
types of host galaxies.  In particular, galaxies with high optical luminosities
are absent.  Using the expression given in \S~12.5 of Barger et al.\ (2002)
and the information listed in Table~1, we find that 14 (78\%) of the objects
we observed have $B$-band luminosities below $10^{43}$ ergs~s$^{-1}$, and all
have $L_B < 2 \times 10^{43}$ ergs~s$^{-1}$.  In contrast, the majority of the
sources spectroscopically classified by Barger et al.\ have $L_B$ in excess of
$10^{43}$ ergs~s$^{-1}$.  Compared to the objects in our sample, the integrated
light of these high-$L_B$ galaxies could more easily overwhelm the emission
from moderately luminous Seyfert nulcei.  Host-galaxy luminosity may also be
related to the detection of optically dull, X-ray--bright objects at fairly
low redshifts ($z < 0.4$; see Comastri et al.\ 2002b).  The possibility that
some host-galaxy light is excluded in the spectra of these objects is probably
offset by the fact that their host galaxies are quite luminous --- nine of
the ten ``normal'' galaxies in the Comastri et al.\ (2002b) study have $L_B$
greater than a few times $10^{43}$ ergs~s$^{-1}$.

Thus, as we have speculated previously (Moran et al.\ 2001), it {\it is\/}
possible to ``hide'' the true nature of classical type~2 AGNs in their
integrated optical spectra.  Low signal-to-noise ratio and inadequate spectral
coverage due to
redshift effects can hinder the classification process further.  We conclude,
therefore, that the demographics of the distant X-ray galaxy population under
investigation in deep {\it Chandra\/} surveys are more likely to be influenced
by the limitations of ground-based spectroscopic observations than by some
evolution of the population.  The majority of the apparently normal galaxies
that have turned up in these surveys, at least those with flat (presumably
absorbed) X-ray spectra and/or substantial X-ray luminosities, would probably
have optical spectra similar to those of nearby Seyfert~2s if their nuclei
could be isolated spatially.  If so, this would verify that Seyfert~2 galaxies
are the dominant contributors to the hard XRB, as prior research has suggested.
Host galaxy dilution is likely to impact any ground-based spectroscopic survey
of distant galaxies, leading to underestimates of the numbers of narrow-line
AGNs in such surveys.

\acknowledgments
We would like to express our thanks to Bill Holzapfel for generously donating
the August 2001 Lick time that enabled us to get this project underway.  This
work was supported by NASA grant NAG5-3556 to A.~V.~F.

\begin{center}
\begin{deluxetable}{lcccc}
\tablewidth{0pt}
\tablecaption{Seyfert 2 Galaxies Observed}
\tablehead{\colhead{Galaxy} &
           \colhead{$d$ (Mpc)\tablenotemark{a}} &
           \colhead{$m$\tablenotemark{b}} &
           \colhead{Exp.\ (s)\tablenotemark{c}} &
           \colhead{Class.\tablenotemark{d}}}
\startdata
MCG $+$01-27-020         &  46.8  & 14.7 & 120~ & Sy2\\
Mrk 3                    &  54.0  & 13.6 & 60   & Sy2\\
Mrk 1066                 &  48.1  & 13.6 & 82   & Sy2\\
NGC 262                  &  60.1  & 13.9 & 75   & Sy2\\
NGC 591                  &  60.7  & 13.9 & 120~ & Sy2\\
NGC 788                  &  54.4  & 13.1 & 60   & N/e\\
NGC 1358                 &  53.8  & 12.7 & 30   & N/e\\
NGC 1667                 &  60.7  & 13.1 & 60   & ~~N/sb\\
NGC 1685                 &  60.4  & 14.5 & 60   & N/e\\
NGC 2273                 &  28.4  & 12.3 & 72   & N/e\\
NGC 3081                 &  32.5  & 12.8 & 40   & Sy2\\
NGC 3982                 &  27.2  & 11.8 & 36   & ~~N/sb\\
NGC 4117                 &  17.0  & 14.0 & 48   & N/e\\
NGC 5283                 &  41.4  & 14.2 & 90   & Sy2\\  
NGC 5347                 &  36.7  & 13.2 & 60   & N/e\\
NGC 5695                 &  56.4  & 13.8 & 60   & N/e\\
NGC 5929\tablenotemark{e}&  38.5  & 13.0 & 40   & N/e\\
NGC 7672                 &  53.5  & 14.4 & 86   & N/e
\enddata
\tablenotetext{a}{Distances from Tully (1988).  For the more distant 
                  sources ($cz > 3000$ km~s$^{-1}$), we assume
                  $d = cz/H_0$, where $H_0$ = 75 km~s$^{-1}$ Mpc$^{-1}$.}
\tablenotetext{b}{Optical $B$ band magnitude, from the NASA Extragalactic
                  Database (NED).}
\tablenotetext{c}{Effective exposure in the scanned observations.}
\tablenotetext{d}{Classification based on the integrated spectrum.
                  Sy2 = Seyfert 2; N/e = normal galaxy with early-type
                  continuum; N/sb = normal galaxy with starburst emission
                  lines.}
\tablenotetext{e}{Interacting with NGC 5930, which is included in the
                  integrated spectrum.}
\end{deluxetable}
\end{center}

\begin{figure}
\begin{center}
\epsscale{1.0}
\centerline{\plotone{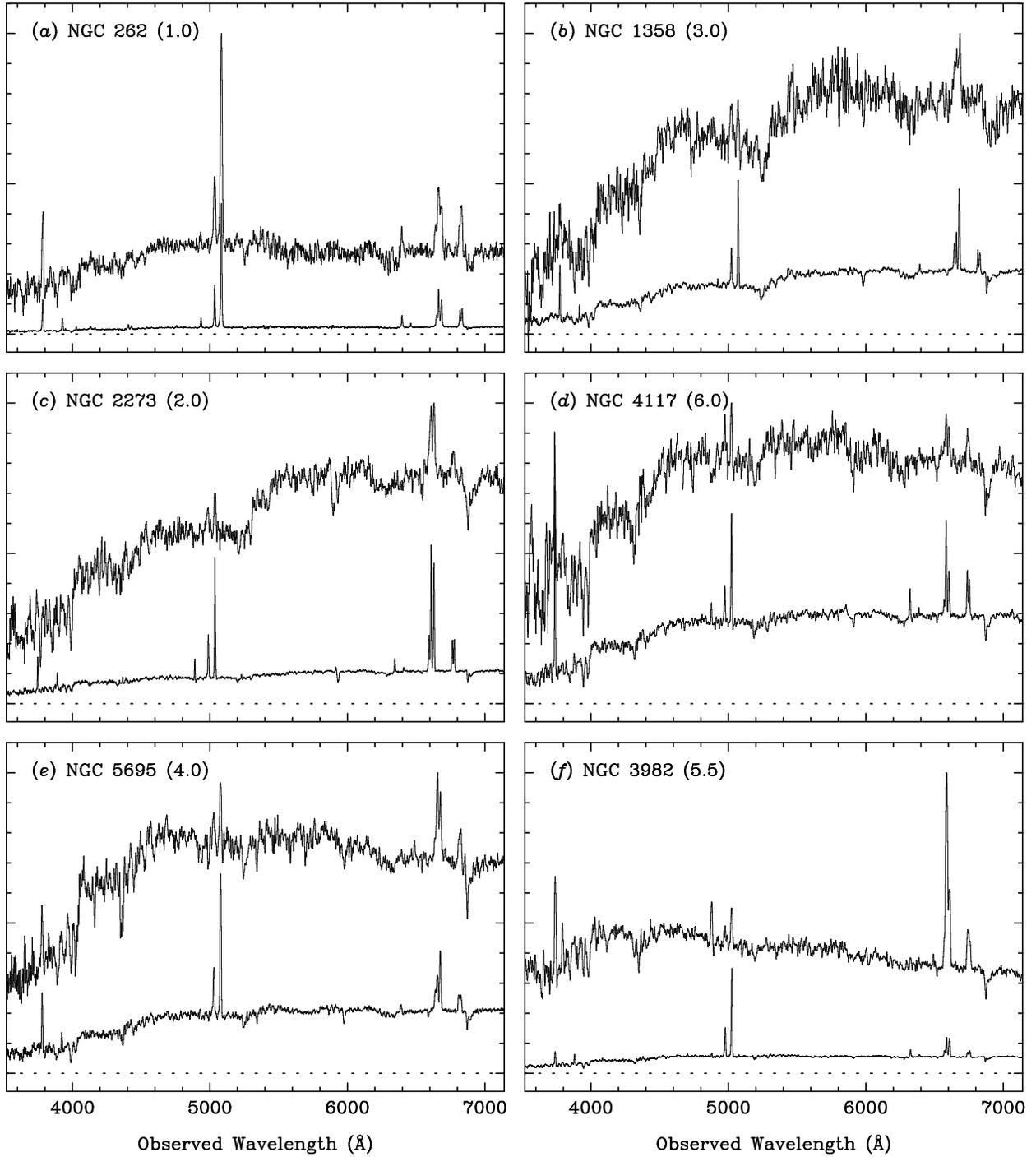}}
%\vskip 0.1truein
\caption{Data for six of the observed Seyfert 2 galaxies.  Both nuclear
(lower trace) and integrated (upper trace) spectra are shown in each panel.
Relative flux densities in units of ergs cm$^{-2}$ s$^{-1}$ \AA $^{-1}$ are
plotted on the ordinate; the dotted line indicates the zero flux level.
For clarity, the nuclear spectrum of each object has been multiplied by a
constant, which appears following the galaxy name.}
\end{center}
\end{figure}

\end{document}